# Research on Optimal Portfolio Based on Multifractal Features

Yong Li

Business School, China University of Political Science and Law, Beijing 100085, China;

yongli@cupl.edu.cn

**Abstract**: Providing optimal portfolio selection for investors has always been one of the hot topics in academia. In view of the traditional portfolio model could not adapt to the actual capital market and can provide erroneous results. This paper innovatively constructs a mean-detrended cross-correlation portfolio model (M-DCCP model), This model is designed to embed detrended cross-correlation between different simultaneously recorded time series in the presence of nonstationary into the reward-risk criterion. We illustrate the model's effectiveness by selected five composite indexes (SSE 50, CSI 300, SSE 500, CSI 1000 and CSI 2000) in China A-share market. The empirical results show that compared with traditional mean-variance portfolio model (M-VP model), the M-DCCP model is more conducive for investors to construct optimal portfolios under the different fluctuation exponent preference and time scales preference, so as to improve portfolio's performance.

## 1. Introduction

Providing optimal portfolio selection for investors has always been one of the hot topics in academia. Since Markowitz proposed the mean-variance portfolio model (M-VP model) in 1952[1], a number of studies have attempted to demonstrate the potential of correlation for the reward-risk criterion, and to find the major determinants of risk diversification and presented some improved portfolio optimization models. For example, mean-absolute deviation models[2, 3], mean-semi-variance model[4, 5], mean-variance-VAR (CVAR) model [6-9] and so on. Although the above researches have greatly enriched modern portfolio theory, but those models all based on the efficient market hypothesis, which take it for granted that the return of financial assets follows the normal distribution. Quantifying cross-correlations between risky assets by covariance based on the Gaussian statistical methods may be not sufficient and can provide erroneous results. Especially in the case, when the actual capital market is the nonlinear dependent structure.

Existing studies have shown that the capital market has multifractal characteristics, and the correlation of different asset prices/return in different fluctuation exponents and different time scales is different[10]. Could the multifractal correlation characteristics of asset prices/return be incorporated into the traditional return-risk criterion? Is it possible to design a novel portfolio model under the realistic background of multifractal characteristics of the capital model to improve the portfolio optimization? This paper tries to answer these questions. The rest of this paper is structured as follows: The second part is literature review; The third part is the model construction, a mean-detrended cross-correlation portfolio model (M-DCCP model) is constructed under the constraints of multifractal characteristics of the capital market; The fourth part is empirical analysis to verify the effectiveness of the M-DCCP model. The last part is the conclusion.

2. Literature Review

Since the 1990s, a large number of anomalies (such as the small-firm effect, January effect, reversals, etc.) appeared in the capital market are recognized. In order to better analyze and interpret those financial anomalies, many researchers try to use multifractal techniques to study the characteristics of capital markets. Kantelhardt [11] proposed multifractal detrended fluctuation analysis (MF-DFA for short), Podobnik and Stanley [12] proposed detrended cross-correlation analysis (DCCA). Then Zhou[13] combined MF-DFA and DCCA to form multifractal detrended cross correlation analysis (MF-DCCA). Empirically, Kim et al. [14] used MF-DFA to study the nonlinear characteristics of South Korea's fund market, and Sukpitak & Hengpunya [15]used MF-DFA to analyze the complexity behavior characteristics of Thailand's capital market. Ma et al. [16]used the DCCA method to study the efficiency and correlation of Chinese stock market before and after the financial crisis. And MF-DCCA method has also been widely used to reveal the multifractal characteristics of financial markets[17, 18], energy market[19], and air pollution fields [20].

Recently, some scholars have tried to apply multifractal techniques to asset portfolio selection. Tilfan et al. [21] proposed the improved M-VP model with the aid of fractal regression to optimize portfolio selection, and carried out the empirical test in four emerging stock markets: China, the Czech Republic, Hungary and Russia. Li et al., [22]constructed a mean-MF-DCCA portfolio model by considering the assets' nonlinearity with specific strategies under the assumption of constant scale preference. Kakinaka [9] examined whether time scale preference of investors improves portfolio performance using the data of S&P500 stock, US Treasury bond, US High-yield corporate bond, and US Investment-grade corporate bond. Although the previous studies have made some useful explorations on the application of multifractals to portfolio selection, but their researches are limited to the portfolio selection of two assets, and accounting for changes in investors' scale preference in response to market conditions. However, it is a well-known fact that both the time scale and fluctuation exponent are important determining factors of the assets' correlation and the portfolio's risk diversification, simple analysis based on time scale may provide biased information. There are still lack of the exploration of multifractal technology used for multi-asset portfolio optimization.

In this paper, multifractal techniques is incorporated into the optimal portfolio selection framework, a M-DCCP model is innovatively proposed by embedding detrend cross-correlation function to substitute covariance into the reward-risk criterion, which fully consider the nonlinear dependent structure of assets under investors' preference for scale pattern and fluctuation exponent.

Compared with existed achievements, the marginal contributions of this paper are as follows: we construct a M-DCCP model by using multifractal detrended cross-correlation function to replace the covariance in M-VP model, which can reflect the realistic background of multifractal correlation characteristics of asset prices/returns, and risk diversification metric fully considers the impact of investors' preference in the multiple fluctuation exponent and multiple time scales. The five composite indexes (SSE 50, CSI 300, SSE 500, CSI 1000 and CSI 2000) in China A-share market is used to test the effectiveness of the M-DCCP model. The empirical results show that compared with the M-VP model, the M-DCCP model is more conducive for investors to construct optimal portfolios under the different fluctuation exponent preference and time scales preference. The M-DCCP model based on multifractal analysis not only enrich the modern portfolio theory, but also improve the performance of investment portfolio.

**3. Model Building**

In order to incorporate the multifractal techniques into the framework of portfolio selection, it is necessary to embed multifractal risk metric into the return-risk criterion. So, we first briefly demonstrate the traditional M-VP model, then find the risk measure based on multifractal techniques and construct the M-DCCP model, and derive its analytic solutions.

**3.1** The M-VP model

The traditional M-VP model can be constructed from the following procedure: Assume that investors plan to use N risky assets $\{Y_i\}_{i=1}^{N}$ to construct a multi-asset portfolios portfolio P, let $r_i$ is the expected return of asset $Y_i$, $Cov(r_i,r_j)$ represents the covariance of assets $Y_i$ and $Y_j$, $\omega_i$ is the weight of the asset i. Obviously, the risk of asset $Y_i$ is the variance $Var(r_i)$, that is, its own covariance $Cov(r_i,r_i) = Var(r_i)$. Based on the assumption of rational behavior, investors choose the optimal portfolios that give the highest possible expected return for a given risk level or the lowest risk for a desired expected return. Then, we can construct a portfolio model by maximizing the single objective function *Sp*:

$$\text{Max } S_P = \frac{E(r_P) - r_F}{Var(r_P)} = \frac{\sum_{i=1}^{N} \omega_i E(r_i) - r_F}{\sum_{i=1}^{N} \omega_i^2 Var(r_i) + 2\sum_{i,j=1, i \neq j}^{N} \omega_i \omega_j Cov(r_i, r_j)}$$

$$\text{s.t.} \begin{cases} \sum_{i=1}^{N} \omega_i = 1 \\ \omega_i \geq 0 \ (i = 1, 2, \cdots N) \end{cases} \quad (1)$$

To think investors require optimize portfolio P to achieve the minimum risk $Var(r_P)$ under a given return u, it is needed to solve $\omega_i$ through maximizing the objective function under constraints, as shown in equation (2).

$$\begin{cases} \text{Max } S_P = \dfrac{\sum_{i=1}^{N}\omega_i E(r_i)-r_F}{\sum_{i=1}^{N}\omega_i^2 Var(r_i)+2\sum_{i=1}^{N}\sum_{j=1,i\neq j}^{i}\omega_i\omega_j Cov(r_i,r_j)} \\ s.t. \begin{cases} \sum_{i=1}^{N}\omega_i = 1 \\ \sum_{i=1}^{N}\omega_i E(r_i) = u \end{cases} \end{cases} \quad (2)$$

The analytical solution of $\omega_i$ could be obtained by Lagrange multiplier method. In order to simplify the expression, the expected return $E(r_i)$ of asset i is represented by $r_i$. Solving equation (2), the optimal weight $\omega_i$ of asset $Y_i$ is obtained as shown in equation (3), and the expected return of portfolio P can be calculated by equation (4).

$$\omega_i = \dfrac{2u\sum_{j=1}^{N}Cov(r_i,r_j)\left(r_j\sum_{i,j=1}^{N}Cov(r_i,r_j)-\sum_{i,j=1}^{N}r_jCov(r_i,r_j)\right)+2\sum_{j=1}^{N}Cov(r_i,r_j)\left(\sum_{i,j=1}^{N}r_ir_jCov(r_i,r_j)-r_j\sum_{i,j=1}^{N}r_jCov(r_i,r_j)\right)}{\sum_{i,j=1}^{N}r_ir_jCov(r_i,r_j)\sum_{i,j=1}^{N}Cov(r_i,r_j)-(\sum_{i,j=1}^{N}r_jCov(r_i,r_j))^2}$$

(3)

$$Er_P = \sum_{i=1}^{N}\omega_i E(r_i) = \sum_{i=1}^{N}\omega_i r_i \quad (4)$$

As can be seen from equation (3), the optimal weight of assets i is determined by the expected return of assets and covariance between asset pairs. As a measure of the correlation between assets, covariance directly affects the effect of risk diversification in portfolio. Therefore, for the capital market with multifractal characteristics, if the covariance cannot effectively measure the nonlinear dependency relation between asset prices/returns, it will certainly affect the measure of diversification benefits of the portfolio.

**3.2 Multiscale adaptive multifractal cross-correlation techniques**

Referring to the methods of Shi et al. [17] and Li et al. [22], the specific steps for analyzing the multiscale adaptive multifractal cross-correlation characteristics of multivariate time series are as follows:

Suppose that the return series of two financial assets i and j are $\{r_t^i\}$ and $\{r_t^j\}$, t =t$_1$, t$_2$,... T, T is the length of the return series.

Step 1: Constructing the cumulative deviation sequence

$$I_z = \sum_{t=1}^{z}(r_t^i - \bar{\imath}),$$
$$J_z = \sum_{t=1}^{z}(r_t^j - \bar{\jmath}), \quad (5)$$

Here $\bar{\imath}$、$\bar{\jmath}$ is the mean of return series $\{r_t^i\}$ and $\{r_t^j\}$ respectively, z=t$_1$, t$_2$, …,

T.

Step 2: Using the adjusted moving average method to fit the $I_z$ and $J_z$ sequences as follows:

$$\tilde{I}_k = \frac{1}{l}(I_k + I_{k-1} + I_{k-2} + \cdots + I_{k-l}) \quad ,$$
$$\tilde{J}_k = \frac{1}{l}(J_k + J_{k-1} + J_{k-2} + \cdots + J_{k-l}) \quad , \tag{6}$$

Here $l=\text{int}(T/\tau)$, $\tau \leqslant k \leq T$, and the τ value can adjust the moving window width adaptively according to the time series length, and the fitting sequences $\{\tilde{I}_k\}$ and $\{\tilde{J}_k\}$ are obtained.

Step 3: Dividing the cumulative deviation sequence $\{I_z\}$ and $\{J_z\}\}$ and the new sequence $\{\tilde{I}_k\}$ and $\{\tilde{J}_k\}$ into d adjacent non-overlapping boxes of length s, i.e. $d=\text{int}(T/s)$. When the series length may not be an integer multiple of s, in order to make full use of all the data information, we repartition the data from the end of the series. Thus, for a given time scale s, a total of 8d subseries can be gotten for the four series.

By calculating the residuals between each subsequence and the fitting subsequence, the detrended cross-correlation function of each segment v (box v) of the scale s is obtained:

$$F_v(s) = \frac{1}{s}\sum_{k=1}^{s}(I_{vk} - \tilde{I}_{vk}) \cdot (J_{vk} - \tilde{J}_{vk}) \quad , \tag{7}$$

Step 4: Calculating the q-order detrended cross-correlation function for the two return series:

$$\begin{cases} F_{ij}(q,s) = \{\frac{1}{2d}\sum_{v=1}^{2d}[F_v(s)]^{\frac{q}{2}}\}^{\frac{1}{q}}, & q \neq 0, \\ F_{ij}(q,s) = \exp\{\frac{1}{4d}\sum_{v=1}^{2d}ln[F_v(s)]\}, & q = 0. \end{cases} \tag{8}$$

In equation (8), [·] represents the absolute value. It can guarantee the formula valid by correcting any negative numbers that may occur when 1/q power is applied. The detrended cross-correlation functions $F_{ij}(q,s)$ is defined at multifractal order q and detrending scale s. The exponent q acts as a filter. When q>2, the boxes with large fluctuations contribute the most to $F_{ij}(q,s)$, when q ≤ 2, the boxes with relatively small values dominate $F_{ij}(q,s)$. Thus, different q describes the effect of different degrees of fluctuation on $F_{ij}(q,s)$.

Step 5: Analyzing the scaling behavior of the detrended cross-correlation function $F_{ij}(q,s)$ by graphing the $\log F_{ij}(q,s) - \log s$ for each given fluctuation exponent q.

If the two series $\{r_t^i\}$ and $\{r_t^j\}$ exist a long-range power-law interaction, then the following formula is true:

$$F_{ij}(q,s) \propto s^{H_{ij}(q)} ,  \qquad (9)$$

Here, $H_{ij}(q)$ is called the generalized scale index and it describes the power-law correlation of two time series.

The above multifractal analysis shows that the cross-correlation function $F_{ij}(q,s)$ described the nonlinear dependent pattern between assets will be different under the joint action of different fluctuation exponent q and different time scales s, so it can measure the risk diversification effect of the asset portfolio more comprehensively without assuming the distribution of asset returns. Because the traditional covariance based on Gaussian statistical method cannot capture the difference stemmed from the changes in q and s, there exist some defects for it to measure the risk diversification of asset portfolios. Therefore, it is reasonable and feasible to replace traditional covariance with $F_{ij}(q,s)$ in reward-risk criterion.

### 3.3 Building the M-DCCP model

In order to overcome the deficiency of the traditional M-VP model not capturing the nonlinear dependent structure between risky assets, the M-DCCP model is constructed by embedding the multifractal detrend cross-correlation function $F_{ij}(q,s)$ into the framework of portfolio selection. Let's still assume that portfolio P consists of N risky assets, $r_i$ represents the expected return rate of asset i, $r_F$ represents the risk-free return rate, and $\omega_i(q, s)$ represents the weight of asset i. Obviously, the expected return rate $Er_P(q, s)$ of portfolio P meets the equation (10):

$$Er_P(q,s) = \sum_{i=1}^{N} \omega_i(q,s) r_i \qquad (10)$$

Considering $F_{ij}(q, s)$ as a measure of correlation would reveal the dependence relationship between two risky assets i and j in different fluctuation exponent q and different time scale s, and embedding the detrended cross-correlation function $F_{ij}(q, s)$ into the return-risk criterion, the risk metric $Var\, r_P(q, s)$ of portfolio P can be expressed by equation (11):

$$Var\, r_P(q,s) = \sum_{i=1}^{N} \omega_i^2(q,s) F_{ii}(q,s) + 2 \sum_{i,j=1, i \neq j}^{N} \omega_i(q,s) \omega_j(q,s) F_{ij}(q,s)$$

$$(11)$$

The M-DCCP model can be expressed as follows:

$$\begin{cases} \text{Max } S_P(q,s) = \dfrac{\sum_{i=1}^{N} \omega_i(q,s) r_i - r_F}{\sum_{i=1}^{N} \omega_i^2(q,s) F_{ii}(q,s) + 2\sum_{i,j=1, i\neq j}^{N} \omega_i(q,s)\omega_j(q,s) F_{ij}(q,s)} \\ s.t. \begin{cases} \sum_{i=1}^{N} \omega_i(q,s) = 1 \\ \sum_{i=1}^{N} \omega_i(q,s) r_i = u \end{cases} \end{cases} \quad (12)$$

The analytical solution of the model is equivalent to solving the values of $\omega_i(q,s)$ by maximizing objective function $S_P(q, s)$. Solving equation (12), the analytical solution of $\omega_i(q, s)$ can be obtained as equation (13).

$$\omega_i(q,s) =$$
$$\dfrac{2u\sum_{j=1}^{N} F_{ij}(q,s)\left(r_j \sum_{i,j=1}^{N} F_{ij}(q,s) - \sum_{i,j=1}^{N} r_j F_{ij}(q,s)\right) + 2\sum_{j=1}^{N} F_{ij}(q,s)\left(\sum_{i,j=1}^{N} r_i r_j F_{ij}(q,s) - r_j \sum_{i,j=1}^{N} r_j F_{ij}(q,s)\right)}{\sum_{i,j=1}^{N} r_i r_j F_{ij}(q,s) \sum_{i,j=1}^{N} F_{ij}(q,s) - \left(\sum_{i,j=1}^{N} r_j F_{ij}(q,s)\right)^2}$$
(13)

The M-DCCP model provides the optimal weight of constituent assets based on the full use of the fluctuation information of the market by adjusting the parameters q and s, so as to meet the diversified needs of investors in different trading cycles.

In the capital market with multifractal characteristics, investors are heterogeneous in their expectations and they have different investment horizons. In order to achieve the optimal risk diversification benefits of portfolios, it is necessary to construct the asset portfolio under different fluctuation degrees and time scales for different investment levels. Given a multiple time scale set $S\subseteq\{s_{min}, s_{max}\}$, where $s_{min}$ and $s_{max}$ are the smallest and largest elements of the set S respectively, and the fluctuation exponent set $Q\subseteq\{q_{min}, q_{max}\}$, $q_{min}$ and $q_{max}$ are the smallest and largest elements of the set Q, respectively, the optimal weight of the constituent assets $\omega_i(Q,S)$ is:

$$\omega_i(Q,S) = \sum_{s\in S}^{q\in Q} \alpha(q,s)\,\omega_i(q,s) \quad (14)$$

and the expected portfolio return is:

$$Er_P(q,s) = \sum_{i=1}^{N} \omega_i(Q,S) r_i = \sum_{i=1}^{N} \sum_{s\in S}^{q\in Q} \alpha(q,s)\,\omega_i(q,s)\,r_i \quad (15)$$

Where $\alpha(q, s)$ is the relative preference degree of investors for time scale ($s\in S$) and fluctuation range ($q\in Q$). $\alpha(q, s)\in[0,1]$, and $\sum_{s\in S}^{q\in Q} \alpha(q,s) = 1$.

$\alpha(q, s)$ can be adjusted based on the choice of q and s. In view of the heterogeneity of investors, $\alpha(q, s)$ has an infinite possible values for the preference combination of q and s, so it is not practical to list them all one by one. A feasible method can be adopted is to categorize investors' preference for s and q. In reference to Li (2021), investors' preference for s and q is divided into nine categories: ①There is no obvious preference for s and q (C-I); ②Preference for short time scales but no obvious preference for fluctuation exponent (C-II); ③Preference for long time scales but no obvious preference for fluctuation exponent (C-III); ④Preference for large fluctuation exponents but no obvious preference for time scale (C-IV); ⑤Preference for small fluctuation exponents but no obvious preference for time scale (C-V); ⑥Preference for long time scales and large fluctuation exponents (C-VI); ⑦Preference for long time scales and small fluctuation exponents (C-VII); ⑧Preference for short time scales and large fluctuation exponents (C-VIII); ⑨Preference for short time scales and small fluctuation exponents (C-VIIII). The assignment to α(q,s) could base on the investor's acceptance range of time scales and fluctuation exponents. For example, if the investor have no obvious preference for s and q, we could set:

$$\alpha(q,s) = \frac{1}{(s_{max}-s_{min}+1)(q_{max}-q_{min}+1)}, \quad (16)$$

It can be seen that the M-DCCP model uses the detrend cross-correlation function $F_{ij}(q, s)$ to capture the dynamic nonlinear characteristics of constituent assets, which make it adapt to the realistic background of multifractal fluctuation characteristics and multifractal correlation characteristics of asset prices/returns. Moreover, $F_{ij}(q, s)$ can quantify the level of cross-correlations on investors' preference for multiple time scales and multiple fluctuation exponents, which expresses closer to the realistic background of the capital market with multifractal characteristics than the M-VP model.

**4. Effectiveness test of M-DCCP model**

4.1 Data

We use five indices in China's A-share stock market, namely Shanghai Stock Exchange 50 (SSE 50), China Securities 300 Index (CSI 300), China Securities 500 Index (CSI 500), China Securities 1000 Index (CSI 1000) and China Securities 2000 Index (CSI 2000) as samples to analyze the performance differences between the M-DCCP model and the M-VP model, to verify the effectiveness of the M-DCCP model.

Comparing SSE 50, CSI 300, CSI 500, CSI 1000 and CSI 2000, each has its own connotations and represent a specific market type. The SSE 50 consists of the 50 largest stocks with the most liquidity in the Shanghai Stock market, it is designed to reflect the overall performance of the most market influential and high-quality large-cap companies. The CSI 300 consists of the 300 stocks in Shanghai and Shenzhen Stock Exchange, and is an authoritative indicator of the overall stock price performance of the listed companies. The CSI 500 is designed to reflect comprehensively the stock price performance of small and medium-sized companies in China's A-share market. The CSI 1000 index is a comprehensive reflection of the overall state of small-cap stocks. The CSI 2000 is designed to reflect the market performance of smaller securities. The constituent stocks of each index do not overlap with one another. The sample period is from January 1, 2015 to December 31, 2023. And this paper takes each year as a sub-period to fully observe the multifractal properties of China stock market. Data is collected from Wind databases.

Table 1 Descriptive statistics of the return rate of the five indices in the sample period

| Index | Mean | Median | Maximum | Minimum | Std. Dev. | Skewness | Kurtosis |
|---|---|---|---|---|---|---|---|
| SSE 50 | 0.0065 | 0.0030 | 0.3485 | -0.1850 | 0.0643 | 1.2470 | 9.0455 |
| CSI 300 | 0.0063 | 0.0030 | 0.2581 | -0.2104 | 0.0625 | 0.5486 | 6.5734 |
| CSI 500 | 0.0060 | 0.0059 | 0.2759 | -0.2820 | 0.0749 | -0.1187 | 6.9759 |
| CSI 1000 | 0.0057 | 0.0033 | 0.3189 | -0.3042 | 0.0879 | 0.0384 | 6.1446 |
| CSI 2000 | 0.0104 | 0.0085 | 0.3682 | -0.2879 | 0.0943 | 0.2891 | 6.3497 |

As can be seen from Table 1, the distributions of the return rate series data of the five indices are asymmetrical with one tail extending further than the other. Moreover, the kurtosis of the five return rate series is significantly greater than the zero of normal distribution, which exhibits the peak characteristics. The sharp-peak and fat-tail characteristics of the return rate series distributions indicate that the indices may have multifractal fluctuation characteristics.

**4.2 Multifractal detrend cross-correlation analysis of the five indices**

In order to reveal the correlation between the constituent stocks in the five indexes, A pair of constituent stocks are randomly selected from each index for multifractal feature analysis. A total of five stock pairs are shown for their generalized scale index $H_{ij}(q, s)$ Vs. q in Figure 1. (Due to the large number of the pairwise composition of each index, SSE 50 (1225 pairs), CSI 300 (44850), CSI 500 (124750), CSI 1000 (499500) and CSI 2000 (1999000), it is cumbersome to display all of them. To be concise, here, only one pair of constituent stock randomly selected from each index are demonstrated, and other constituent stock combinations have similar multifractal correlation characteristics.)

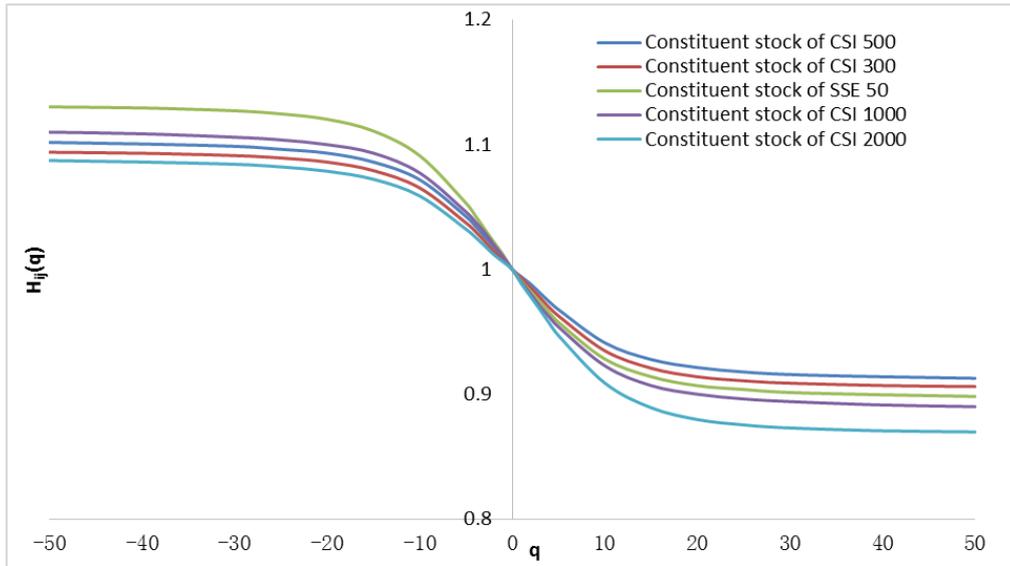

Figure 1. Multifractal scale properties of five constituent stock pairs

Figure 1 shows that the generalized scale index $H_{ij}(q, s)$ for all the five constituent stock pairs changes with q, and the relation shows a monotonically decreasing trend, which indicates that the China A-shares market does have obvious multifractal correlation characteristics. Moreover, the generalized scale index appears $H_{ij}$ (q < 0)≥$H_{ij}$ (q > 0) and $H_{ij}$ (q = 2)≥0, which means that the redundant trend could be detrended well, and the obtained multifractal cross-correlation between constituent stocks is reliable. Figure 2 shows the relationship of $logF_{ij}(q, s)$ vs. logs at given q.

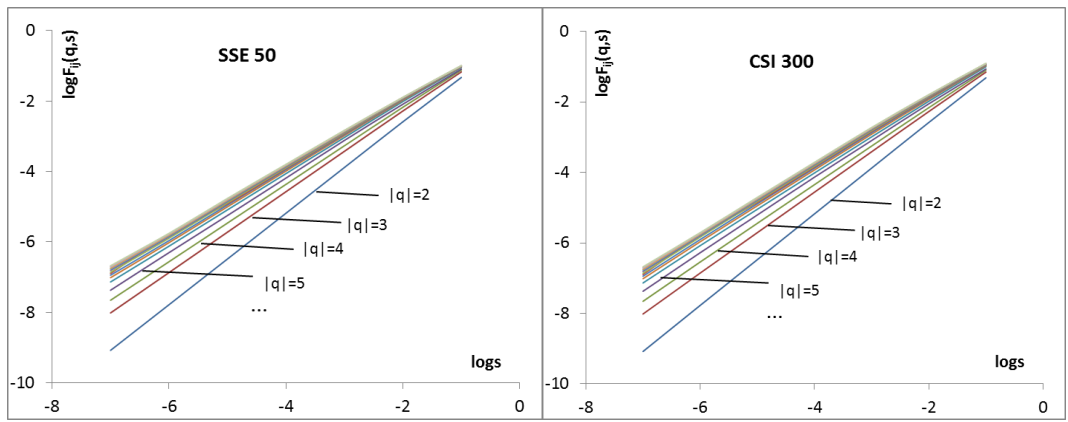

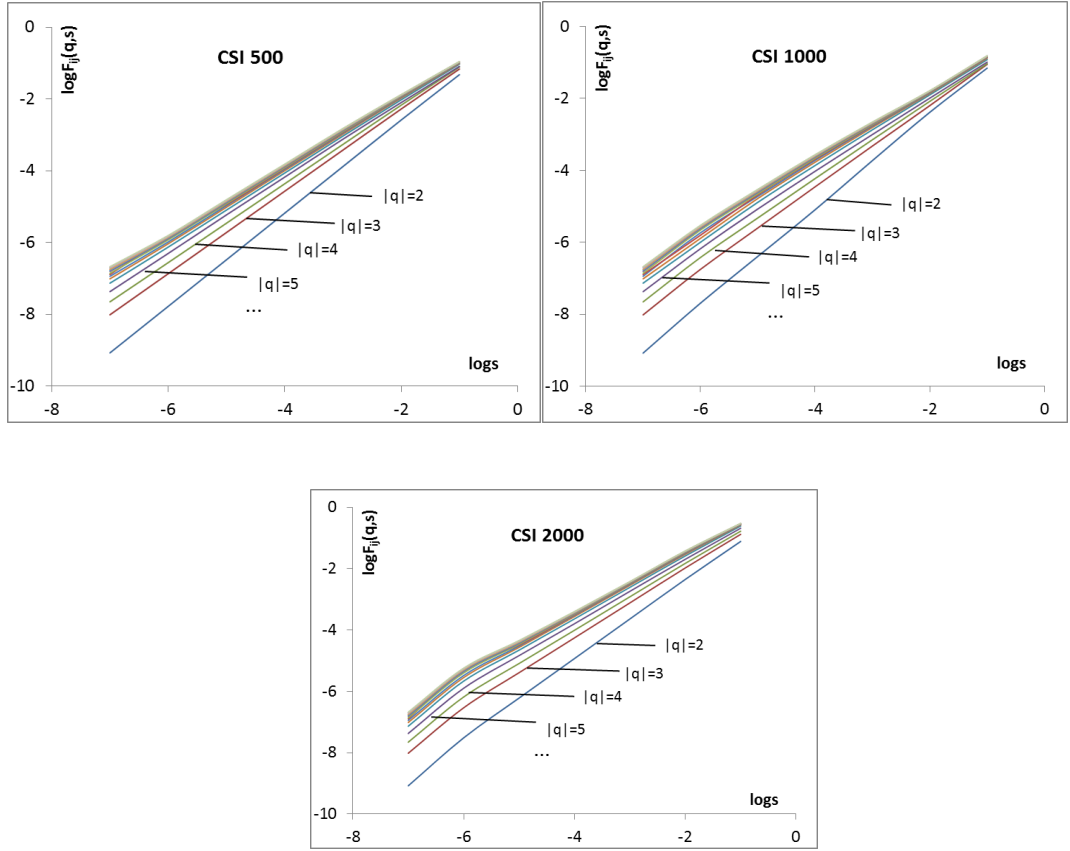

Figure 2. The $\log F_{ij}(q, s)$ vs. logs relationship for 5 indices

As can be seen from Figure 2, the relationship between $\log F_{ij}(q, s)$ and logs can be fitted well and approximately a straight line at given q, which indicate the existence of power-law cross-correlations between $F_{ij}(q, s)$ and s. That once again verifies the existence of multifractal features in the A-share market.

### 4.3 The effectiveness of the M-DCCP model

The method to test the effectiveness of the M-DCCP model is to see whether the benefits of the portfolios can be improved comparing with the M-VP model. From Figure 1 we can detect that when $|q| > 20$, $\log F_{ij}(q, s)$ tends to be stable with respect to q. So, the value of q actually range from -20 to 20 for the A-share market. Additionally, since there are around 242 trading days in a year for China's stock market, the value of s can be set between 3-60 for convenience. That is, $Q \subseteq \{-20, ... , 20\}$ and $S \subseteq \{3,... 60\}$. For SSE 50, CSI 300, CSI 500, CSI 1000 and CSI 2000, each serves as a portfolio of its corresponding constituent stocks, there are 45 stock portfolios under each of the two portfolio models with 9 sample subperiod.

In order to preliminatively judge the superiority of the M-DCCP model, we assume that investors have no obvious preference for s and q, that is, investors have C-I preference, where α (q, s) = $2378^{-1}$, given u = 0.05, the expected return rate of 45 stock portfolios can be calculated by equation (15) for the M-DCCP model and equation (4) for the M-VP model respectively.

Figure 3 shows the expected return rates under the two portfolio models and actual returns. The actual return rate refers to the return rate of the composite index obtained by weighting the market value of the constituent stock, that is, it is Paasche index return.

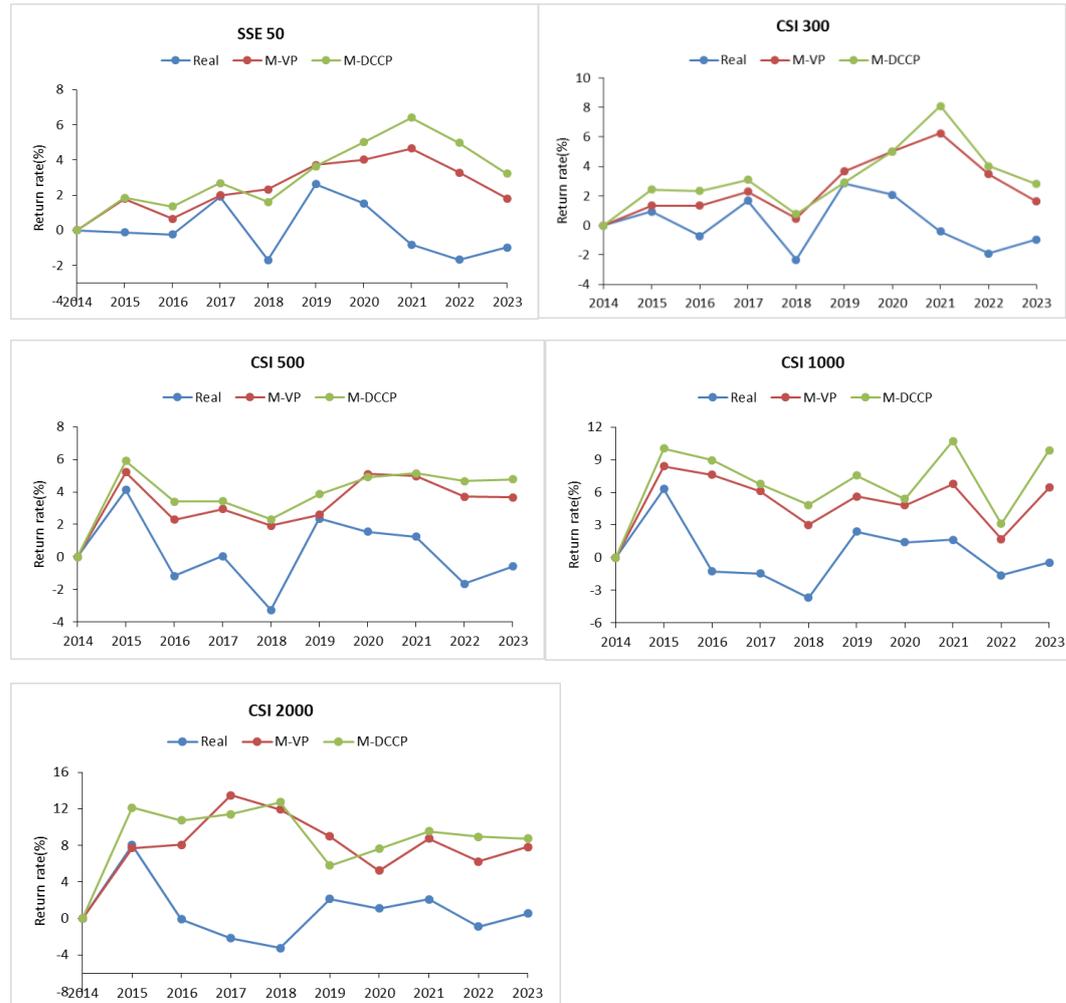

**Figure 3. Expected return rates of five indices under different portfolio models**

Figure 3 shows that the expected return rates of the M-DCCP model and the M-VP model are significantly higher than the actual return rates for all subperiod. The expected return rates of the M-DCCP model are higher than those of the M-VP model except for a few cases, e.g., SSE 50 in 2018, CSI 300 in 2019, CSI 300 in 2020, CSI 2000 in 2017 and 2019. Similar results could be obtained when investors have different preferences for time scales and fluctuation exponents, which are not presented to save the space.

In order to analyze the effectiveness of the M-DCCP model more comprehensively and deeply, the cumulative return rates are calculated under the two model within the nine sample subperiods. Table 2 shows in the three conditions where the vested return rate is set to 0.05, 0.15 and 0.30 respectively, the cumulative expected return rate of the five indices under nine preference categories for the M-DCCP model, compared with those for the M-VP model.

Table 2 Cumulative expected return rates of the M-DCCP and the M-VP model

| u | P | M-VP | M-DCCP | | | | | | | | |
|---|---|---|---|---|---|---|---|---|---|---|---|
| | | | C-I | C-II | C-III | C-IV | C-V | C-VI | C-VII | C-VIII | C-VIIII |
| 0.05 | SSE 50 | 2.957 | 3.638 | 4.340 | 2.935 | 2.858 | 4.418 | 4.001 | 1.714 | 4.680 | 4.155 |
| | CSI 300 | 3.024 | 5.021 | 7.611 | 2.431 | 4.934 | 5.108 | 6.685 | 3.183 | 8.537 | 1.680 |
| | CSI 500 | 3.096 | 5.599 | 6.124 | 3.073 | 3.425 | 4.773 | 4.863 | 2.014 | 3.385 | 2.161 |
| | CSI 1000 | 5.165 | 13.623 | 16.887 | 10.359 | 17.243 | 10.003 | 18.391 | 16.094 | 15.382 | 4.625 |
| | CSI 2000 | 6.020 | 12.492 | 20.698 | 6.286 | 18.964 | 5.234 | 31.015 | 6.912 | 5.661 | 10.380 |
| 0.15 | SSE 50 | 3.306 | 1.845 | 2.339 | 6.030 | 6.223 | 9.914 | 12.395 | 2.052 | 7.716 | 12.111 |
| | CSI 300 | 4.429 | 5.346 | 5.346 | 5.346 | 6.288 | 3.403 | 5.611 | 5.965 | 3.081 | 3.726 |
| | CSI 500 | 4.245 | 10.690 | 12.294 | 9.087 | 11.950 | 9.431 | 11.119 | 12.780 | 13.468 | 5.394 |
| | CSI 1000 | 8.462 | 12.948 | 14.462 | 11.435 | 10.295 | 11.261 | 11.394 | 9.197 | 17.530 | 13.674 |
| | CSI 2000 | 8.681 | 29.641 | 40.672 | 18.609 | 11.590 | 70.872 | 8.618 | 14.563 | 89.962 | 51.781 |
| 0.30 | SSE 50 | 6.902 | 7.010 | 7.012 | 7.008 | 7.105 | 6.815 | 6.807 | 7.403 | 7.217 | 6.613 |
| | CSI 300 | 7.118 | 6.416 | 9.255 | 12.086 | 11.972 | 9.141 | 9.775 | 30.719 | 11.734 | 8.547 |
| | CSI 500 | 8.191 | 5.980 | 13.974 | 12.014 | 10.701 | 12.661 | 7.701 | 9.104 | 6.247 | 6.925 |
| | CSI 1000 | 9.133 | 57.226 | 15.640 | 98.812 | 74.665 | 39.786 | 20.452 | 128.87 | 10.828 | 68.744 |
| | CSI 2000 | 16.079 | 261.62 | 287.47 | 235.77 | 180.93 | 242.31 | 103.02 | 158.83 | 271.91 | 212.71 |

As can be seen from Table 2, when the vested return is set to 0.05, under the preferences of C-I, C-II, and C-VI, the cumulative return rates of the five indices for the M-DCCP model are all greater than those for the M-VP model; Under the case of C-IV, C-V and C-VIII preference, only one index's cumulative return rate for the M-DCCP model is lower than that for the M-VP model, that is SSE 50, CSI 2000 and CSI 2000 respectively; Under the preferences of C-VII, the cumulative return rates of SSE 50 and CSI 500 for the M-DCCP model is lower. Under the C-III and C-VIIII preference, there are three indices' cumulative return rates for the M-DCCP model being smaller than those for the M-VP model, namely SSE 50, CSI 300, CSI 500 and CSI 300, CSI 500, CSI 1000 respectively. So, given u=0.05, there are totally 35 indices case whose cumulative return rates for the M-DCCP model are greater than those for the M-VP model, accounting for 77.78%. When the vested return rate is set to 0.15, under the preferences of C-II, C-III and C-IV, the cumulative return rates of the five indices for the M-DCCP model all are higher than those for the M-VP model; Under the preferences of C-I, C-V, C-VI, C-VII, C-VIII and C-VIIII, each situation there is only one index's cumulative return rate is inferior to that for the M-VP model, they are SSE 50, SSE 300, SSE 50, SSE 300 and SSE 300 respectively. Then, given u=0.15, there are totally 39 indices case whose cumulative return rates for the M-DCCP model are greater, accounting for 86.67%. When the vested return rate is set to 0.30, the cumulative return rates of the M-DCCP model are higher, except for SSE 50 under the C-V, C-VI and C-VIIII preference, the CSI 300 under the C-I preference, and the CSI 500 under the C-I, C-VI, C-VIII and C-VIIII preference. Thus, given u=0.30, there are totally 37 indices case whose cumulative return rates for the M-DCCP model are greater than those for the M-VP model, accounting for 82.22%. Therefore, from the perspective of multi-period investment return rates, the M-DCCP model is verified to be generally better than the M-VP model.

Table 2 verifies the effectiveness of the M-DCCP model from the perspective of cumulative return rates. Considering that the cumulative return rate is a performance indicator without risk adjustment, we further use the cumulative return rate of per unit risk as the performance indicator to compare the two models. The results are shown in Table 3.

Table 3 Risk adjusted expected return rates of the M-DCCP and the M-VP model

| u | P | M-VP | M-DCCP | | | | | | | | |
|---|---|---|---|---|---|---|---|---|---|---|---|
| | | | C-I | C-II | C-III | C-IV | C-V | C-VI | C-VII | C-VIII | C-VIIII |
| 0.05 | SSE 50 | 1.706 | 1.867 | 1.853 | 1.825 | 1.641 | 1.733 | 1.930 | 1.893 | 1.766 | 1.589 |
| | CSI 300 | 0.914 | 1.926 | 1.995 | 1.715 | 1.951 | 1.894 | 1.979 | 1.893 | 2.006 | 1.382 |
| | CSI 500 | 1.212 | 1.512 | 2.140 | 0.906 | 1.450 | 1.550 | 2.156 | 0.981 | 2.120 | 1.007 |
| | CSI 1000 | 1.495 | 2.931 | 2.436 | 3.573 | 2.162 | 3.560 | 2.483 | 1.475 | 2.326 | 1.419 |
| | CSI 2000 | 1.824 | 2.022 | 1.906 | 2.854 | 2.056 | 1.916 | 1.933 | 2.870 | 1.692 | 2.681 |
| 0.15 | SSE 50 | 1.293 | 1.623 | 1.502 | 1.898 | 1.959 | 2.680 | 2.166 | 1.036 | 1.945 | 2.367 |
| | CSI 300 | 1.147 | 1.992 | 1.847 | 1.126 | 1.527 | 1.294 | 1.691 | 1.334 | 1.043 | 1.747 |
| | CSI 500 | 1.626 | 5.129 | 20.798 | 2.505 | 3.868 | 8.669 | 16.181 | 2.311 | 26.106 | 3.121 |
| | CSI 1000 | 1.758 | 2.945 | 3.274 | 2.506 | 2.556 | 2.315 | 2.650 | 2.350 | 3.853 | 2.611 |
| | CSI 2000 | 1.935 | 1.841 | 1.923 | 2.647 | 2.305 | 1.963 | 2.581 | 2.084 | 1.978 | 1.846 |
| 0.30 | SSE 50 | 1.013 | 1.733 | 1.601 | 1.886 | 1.742 | 1.724 | 1.561 | 1.939 | 1.640 | 1.827 |
| | CSI 300 | 1.246 | 1.142 | 2.045 | 1.827 | 1.867 | 2.155 | 1.807 | 1.887 | 2.162 | 2.141 |
| | CSI 500 | 1.501 | 2.111 | 3.830 | 1.984 | 1.147 | 4.651 | 1.959 | 2.000 | 3.151 | 1.479 |
| | CSI 1000 | 1.822 | 1.910 | 1.855 | 1.918 | 1.903 | 1.922 | 1.635 | 1.915 | 1.915 | 1.922 |
| | CSI 2000 | 1.878 | 1.885 | 1.870 | 1.890 | 1.884 | 1.885 | 1.211 | 1.891 | 1.881 | 1.890 |

Table 3 illustrates, when the vested return rate is set to 0.05, the cumulative return rates of per unit risk of the M-DCCP model are higher than those of the M-VP model, except for SSE 50 under C-IV and C-VIIII preference, CSI 500 under C-III, C-VII and C-VIIII preference, CSI 1000 under C-VIIII preference, and CSI 2000 under C-VIII preference; When the vested return rate is set to 0.15, the performances of the M-DCCP model are superior to those of the M-VP model, except for SSE 50 under C-VII preference, CSI 300 under C-III and C-VIII preference, CSI 2000 under C-I and C-VIIII preference; When the vested return rate is set to 0.30, the performances of the M-DCCP model are superior to those of the M-VP model, except for CSI 300 under C-I preference, CSI 500 under C-IV and C-VIIII preference, CSI 1000 under C-VI preference, and CSI 2000 under C-II and C-VI preference. With the vested return rate equals 0.05, 0.15 and 0.30, there are 38, 40 and 39 index cases respectively whose risk adjusted cumulative return rates of the M-DCCP model are greater than those of the M-VP model, accounting for 84.44%, 88.89% and 86.67% respectively. It is obviously found that the M-DCCP model could still provide higher performance than the M-VP model under the risk-adjusted performance measure, indicating that the M-DCCP model is effective.

The above analysis shows that, considering multifractal characteristics of the stock market, the M-DCCP model is generally superior to the M-VP model, no matter from the perspective of single-period return rate, or multi-period cumulative return rate, and no matter from the perspective of return rate without risk adjustment, or from the perspective of risk-adjusted return rate. The advantages of the M-DCCP model over the M-VP model are more obvious and robust. Therefore, the empirical results support the theoretical analysis and show that the M-DCCP model constructed is effective.

## 5. Conclusions

Considering the realistic background of nonlinear dynamics and similarity features of the capital market, this paper incorporates the multifractal techniques into the framework of the return-risk criterion, constructs a M-DCCP model to optimize portfolio selection, and gives the analytical solution of the model. The setting of the M-DCCP model fully considers the non-linear dependent correlation in multiple fluctuation exponents and time scales of the risky assets, which make it closely adapt to the realistic background of multifractal characteristics of the capital market and can overcome the shortcomings of M-VP model to some certain extent. The empirical analysis of the SSE 50, CSI 300, CSI 500, CSI 1000 and CSI 2000 indices of China A-share market shows that the performance of the M-DCCP model is generally superior to that of the M-VP model, indicating that the M-DCCP model is effective. This study successfully realizes the combination of traditional financial theory and multifractal thought, and the M-DCCP model constructed not only enrich modern portfolio theory, but also provides better decision-making reference for investors to improve portfolio performance.

**Funding**: This research was funded by the Fundamental research Funds for the Central University, The scientific research and innovation project of China University of Political Science and Law, grant number: CUPL10824412.

**Conflicts of Interest**: The author declares no conflict of interest.